\documentclass[12pt]{article}
\pdfoutput=1

\usepackage{slashed}
\usepackage{color, verbatim}
\usepackage{latexsym}
\usepackage{amsmath}
\usepackage{graphicx}
\usepackage{arydshln}
\usepackage{cite}
\usepackage{amssymb}
\usepackage{graphicx}
\usepackage{bm}
\usepackage{adjustbox}
\usepackage{booktabs}
\usepackage{multirow}
\usepackage{rotating}

\makeatletter
\@addtoreset{equation}{section}
\g@addto@macro\bfseries{\boldmath}
\makeatother

\newcommand{\ctoprule}{\toprule[0.5mm]}
\newcommand{\cbottomrule}{\bottomrule[0.5mm]}
\newcommand{\cmrule}{\midrule[0.25mm]}

\newcommand{\vtext}[1]{\begin{sideways}\small{#1}\end{sideways}}

\begin{document}
\thispagestyle{empty}
\begin{flushright}
IPPP/19/43 \\
FTUV-19-0524 \\
IFIC/19-28
\end{flushright}
\vspace{0.8cm}

\begin{center}
{\Large\sc Probes of the Standard Model effective field\\[0.5cm] theory extended with a right-handed neutrino}

\vspace{0.8cm}

\textbf{Julien Alcaide$^{\,a,b}$, Shankha Banerjee$^{\,b}$, Mikael Chala$^{\,b,c}$ and Arsenii Titov$^{\,b}$}\\

\vspace{1cm}
{\em {$^a$Departament de F\'isica Te\`orica, Universitat de Val\`encia and IFIC, Universitat de Val\`encia-CSIC,
Dr. Moliner 50, E-46100 Burjassot (Val\`encia), Spain}}\\[0.2cm]
{\em {$^b$Institute for Particle Physics Phenomenology, Department of Physics, Durham University, South Road, 
Durham DH1 3LE, United Kingdom}}\\[0.2cm]
{\em {$^c$CAFPE and Departamento de F\'isica Te\'orica y del Cosmos,
Universidad de Granada, E-18071 Granada, Spain}}
\vspace{0.5cm}
\end{center}

\begin{abstract}
\noindent
If neutrinos are Dirac particles and, as suggested by the so far null LHC results, 
any new physics lies at energies well above the electroweak scale, 
the Standard Model effective field theory has to be extended with operators 
involving the right-handed neutrinos. In this paper, we study this effective 
field theory and set constraints on the different dimension-six interactions. 
To that aim, we use LHC searches for associated production of light (and tau) 
leptons with missing energy, monojet searches, as well as pion and tau decays. 
Our bounds are generally above the TeV for order one couplings. 
One particular exception is given by 
operators involving top quarks. These provide new signals in top decays not yet 
studied at colliders. Thus, we also design an LHC analysis to explore these 
signatures in the $t\overline{t}$ production. Our results are also valid if the 
right-handed neutrinos are Majorana and long-lived.
\end{abstract}

\newpage

\tableofcontents

\section{Introduction}
%
Despite its impressive success in describing particle physics phenomena across a wide range of energies, the Standard Model (SM) is known to be incomplete. The main experimental indication of new physics is the non-vanishing neutrino masses. On the theory side, the hierarchy problem, the flavour puzzle and the unification of couplings among others, also suggest the existence of particles beyond the SM. In light of the null results at the Large Hadron Collider (LHC),  new particles might lie at energies, $\Lambda$, well above the electroweak (EW) scale, established by the Higgs vacuum expectation value (VEV), $v\sim 246$ GeV.

For such scenarios, new physics can be described by the SM effective field theory 
(SMEFT)~\cite{Buchmuller:1985jz}. This theory extends the renormalisable SM 
Lagrangian with a tower of effective operators respecting the SM gauge 
symmetries, but not necessarily the global (accidental) ones. Within this 
framework, neutrinos are predicted to be Majorana, their mass being $m_\nu\sim 
\alpha_5 v^2/\Lambda$ with $\alpha_5$ being the Wilson coefficient of the Weinberg 
operator~\cite{Weinberg:1979sa}. Up to flavour indices, the latter can be considered 
to be the only independent
effective operator of dimension five. At dimension six, this number goes up to 
59~\cite{Grzadkowski:2010es}, assuming baryon number conservation. 
However, if neutrinos are Dirac particles, the SMEFT has to be extended with 
effective operators containing the right-handed (RH) neutrino, $N$. 

This extended EFT, known 
as $\nu$SMEFT, was first considered in Ref.~\cite{delAguila:2008ir}, 
where also a complete set of operators in the $\nu$SMEFT to dimension six was worked out. (See also Ref.~\cite{Aparici:2009fh} for dimension five 
and Ref.~\cite{Bhattacharya:2015vja} for dimension seven.) 
This set was shown to be redundant in Ref.~\cite{Liao:2016qyd}, 
where an actual basis was provided. It is shown in Tab.~\ref{tab:basis} 
for completeness. In this paper, we set constraints on these operators upon 
using searches for one lepton ($e, \mu, \tau$) and missing energy at the LHC, monojet searches at the LHC as well 
as pion and tau decays. For those operators not yet constrained by these 
observables, most of them involving top quarks, we suggest a novel search 
strategy based on a new rare top decay at the LHC.

A comment about cosmological constraints is in order. It is well known that in the absence of any other interactions, sterile neutrinos do not contribute to the number of relativistic species, $N_\text{eff}$. This is no longer true in the $\nu$SMEFT with a low cutoff $\Lambda$. Indeed, the interaction rate for producing RH neutrinos out of the thermal bath behaves as $\Gamma\sim T^5/\Lambda^4$. The latter are in thermal equilibrium at the time of the Big Bang Nucleosynthesis (BBN) provided $\Gamma\gtrsim T_{\text{BBN}}^2/m_P$,
with $m_P$ being the Planck mass and $T_{\text{BBN}}\sim $ MeV; namely if $\Lambda \lesssim 200$ GeV $\sim v$. 
(There is no experimental evidence of the pre 
BBN era~---~some models 
even require reheating right before $T\sim $ 
MeV~\cite{Nakayama:2008wy,Cui:2018rwi}~---~so there are no 
actual 
cosmological
constraints on the $\nu$SMEFT if $N$ decouples before BBN.)

Finally, $N$ could be also a very light Majorana neutrino with mass $m_N$. Even 
in the standard cosmological history, the contribution of such particle to 
$N_\text{eff}$ is below the current limit from Planck, $\Delta 
N_\text{eff}\lesssim 0.3$~\cite{Aghanim:2018eyx}, 
provided $m_N \gtrsim 10$~MeV~\cite{Escudero:2018mvt}.  
Our analysis still applies to this case
if the lifetime of $N$, $\tau_N\sim 256\,\pi^3\Lambda^4/m_N^5$ is 
larger than the scales relevant for 
the experiments considered in this work. 
Namely, for $m_N \lesssim 10^{-3} \Lambda^{4/5}$. 
Thus, for $\Lambda\gtrsim v$, $N$ behaves also as a stable particle if $m_N 
\lesssim 0.1$ GeV. In summary, our study works within the $\nu$SMEFT if the 
cutoff is above the EW scale and provided neutrinos are Dirac particles, 
or Majorana particles with 
$0.01~\mathrm{GeV} \lesssim m_N \lesssim 0.1$~GeV; 
the upper limit being significantly 
larger if the EFT starts being valid only above the TeV.

\begin{table}[t]
\begin{center}
\begin{tabular}{c c l c c l }
\ctoprule
&Operator& Notation& &Operator&Notation\\
\cmidrule{2-3}\cmidrule{5-6}
%
\multirow{3}{*}{\vtext{SF}}&$(\overline{l_L}N)\tilde{H} (H^\dagger H)$&${\cal O}_{lNH}$ (+h.c.)& &&\\
&$(\overline{N}\gamma^\mu N)(H^\dagger i \overleftrightarrow{D_\mu} H)$&${\cal O}_{HN}$&&$(\overline{N}\gamma^\mu e_R)(\tilde{H}^\dagger i D_\mu H)$&${\cal O}_{HNe}$ (+h.c.)\\
&$(\overline{l_L}\sigma_{\mu\nu} N)\tilde{H}B^{\mu\nu}$&${\cal O}_{NB}$ (+h.c.)&&$(\overline{l_L}\sigma_{\mu\nu} N)\sigma_I\tilde{H}W^{I \mu\nu}$&${\cal O}_{NW}$ (+h.c.)\\[0.1cm]
\cmrule
\multirow{3}{*}{\vtext{RRRR}}&$(\overline{N}\gamma_\mu N)(\overline{N}\gamma^\mu N)$&${\cal O}_{NN}$& &&\\
&$(\overline{e_R}\gamma_\mu e_R)(\overline{N}\gamma^\mu N)$&${\cal O}_{eN}$&&$(\overline{u_R}\gamma_\mu u_R)(\overline{N}\gamma^\mu N)$&${\cal O}_{uN}$\\
&$(\overline{d_R}\gamma_\mu d_R)(\overline{N}\gamma^\mu N)$&${\cal O}_{dN}$&&$(\overline{d_R}\gamma_\mu u_R)(\overline{N}\gamma^\mu e_R)$&${\cal O}_{duNe}$ (+h.c.)\\[0.1cm]
\cmrule
\multirow{1}{*}{LLRR}&$(\overline{l_L}\gamma_\mu l_L)(\overline{N}\gamma^\mu N)$&${\cal O}_{lN}$&&$(\overline{q_L}\gamma_\mu q_L)(\overline{N}\gamma^\mu N)$&${\cal O}_{qN}$\\[0.1cm]
\cmrule
\multirow{2}{*}{\vtext{LRRL}}&$(\overline{l_L} N)\epsilon (\overline{l_L}e_R)$&${\cal O}_{lNle}$ (+h.c.)&&$(\overline{l_L} N)\epsilon (\overline{q_L} d_R)$&${\cal O}_{lNqd}$ (+h.c.)\\
&$(\overline{l_L}d_R)\epsilon (\overline{q_L} N)$&${\cal O}_{ldqN}$ (+h.c.)&&$(\overline{q_L}u_R)(\overline{N}l_L)$&${\cal O}_{quNl}$ (+h.c.)\\[0.1cm]
\cbottomrule
\end{tabular}
\caption{Basis of lepton and baryon number conserving dimension-six operators 
containing a 
RH neutrino $N$~\cite{Liao:2016qyd}.
$l_L$ and $q_L$ stand for the left-handed lepton and quark doublets, 
respectively. Likewise, $e_R$ and $u_R$ and $d_R$ stand for the  
right-handed leptons and the up and down quarks, respectively. We use the symbol $H$ to 
denote the Higgs doublet, while $\tilde{H} = \epsilon H^*$, 
where $\epsilon$ is the 
fully antisymmetric tensor in two dimensions. $B_{\mu\nu}$ and 
$W^I_{\mu\nu}$ represent the weak field strength tensors. Flavour indices 
are not shown explicitly.}\label{tab:basis}
\end{center}
\end{table}
%

This article is organised as follows. In section~\ref{sec:formalism} we 
introduce the formalism of the $\nu$SMEFT. 
We discuss which operators contribute to each of the observables considered 
in this work and compute the relevant equations. In section~\ref{sec:fit} we 
compare the obtained results with the experiments, and set global constraints 
on the (four-fermion) operators of the $\nu$SMEFT. We emphasise that some of 
them are actually 
not bounded by current data, and thus develop a new collider search 
designed for the HL-LHC in section~\ref{sec:lhc}. 
We conclude in section~\ref{sec:conclusions}.

\section{Formalism}
\label{sec:formalism}
%
The basis of dimension-six operators containing $N$ and respecting lepton and 
baryon numbers is shown in Tab.~\ref{tab:basis}. In the following subsections, 
we compute different cross sections and decay widths involving  
$N$ in the final state. Therefore, up to corrections proportional to the 
neutrino masses, the $\nu$SMEFT operators mediating the 
corresponding processes do not interfere with their SMEFT counterparts.  As a 
result, any bounds computed on the $\nu$SMEFT Wilson coefficients 
by using solely our equations are conservative. 
Hereafter, we assume all these coefficients to be real. For simplicity, 
we also restrict to only one $N$ family; expressions for the case of several 
$N$ fields can be trivially obtained from our results. Moreover, we explicitly do not 
consider the operators involving flavour violation between the first and 
second fermion families.

We focus mostly on observables sensitive to the four-fermion 
operators in the classes RRRR, LLRR and LRRL, where L (R) denotes left (right) handed fermions. These are typically 
more relevant because they can be generated at tree level in UV completions of 
the SM. Moreover, operators such as $\mathcal{O}_{NB}$ and $\mathcal{O}_{NW}$ 
are very much constrained by measurements of neutrino dipole moments 
when the latter are Dirac~\cite{Canas:2015yoa,Miranda:2019wdy}. In 
the Majorana case they must be suppressed in order for $N$ to be 
long-lived; otherwise it would decay rather promptly into two body 
final states~\cite{Duarte:2015iba,Duarte:2016miz}. 
In such case, the collider signals are very different from the ones 
considered in this article and related works~\cite{Duarte:2014zea,Duarte:2016caz,Caputo:2017pit}; they will be presented elsewhere.
We only include the contribution of the SF operators (see Tab.~\ref{tab:basis}) 
in the relevant equations hereafter for the sake of completeness.

For the observables computed using collider simulations we rely on \texttt{MadGraph v5}~\cite{Alwall:2014hca}, 
\texttt{Pythia v8}~\cite{Sjostrand:2001yu,Sjostrand:2014zea} and 
\texttt{FastJet v3}~\cite{Cacciari:2011ma}, with a model 
containing the interactions in Tab.~\ref{tab:basis} implemented using 
\texttt{FeynRules v2}~\cite{Degrande:2011ua}. 
(For the operator ${\cal O}_{ldqN}$, we use the 
Fierz-transformed version ${\cal O}_{ldqN} = 1/2 (\overline{q_L} 
d_R)\epsilon(\overline{l_L} 
N)+1/8(\overline{q_L}\sigma_{\mu\nu}d_R)\epsilon(\overline{l_L}\sigma^{\mu\nu}
N)$.)

\subsection{Searches for one lepton and missing energy at the LHC}
\label{sec:lNtauN}
%
When restricted to the first (and to a lesser extent, the second) quark 
family, the operators ${\cal O}_{duNe}, {\cal O}_{ldqN}, 
{\cal O}_{lNqd},  {\cal O}_{quNl}$, ${\cal O}_{HNe}$ and 
${\cal O}_{NW}$ contribute to the 
production of $\ell+\slashed{E}_T$ from quarks. In particular,
\begin{align}
\label{eq:pplv}\nonumber
 \frac{d\sigma}{dt} (u\overline{d}\rightarrow \ell_i^+ N) = 
\frac{1}{192\pi\Lambda^4 s^2}&\bigg\lbrace\left[(\alpha_{quNl}^{11i})^2+4 
(\alpha_{duNe}^{11i})^2+(\alpha_{lNqd}^{i11})^2\right] s^2 \\\nonumber
&+\left[4(\alpha_{duNe}^{11i})^2 + (\alpha_{ldqN}^{i11})^2\right] t^2 + 
2\left[4(\alpha_{duNe}^{11i})^2 - 
\alpha_{lNqd}^{i11}\,\alpha_{ldqN}^{i11}\right] st\\
& + 4 (\alpha_{HNe}^i)^2 m_W^4 
\frac{t^2}{s^2} -32(\alpha_{NW}^i)^2 m_W^2\left(\frac{t^2}{s}+t\right)
 \bigg\rbrace~,
\end{align}
%
where $i=1~(2)$ for electron (muon).
The imprint of these operators in searches for $\ell+\slashed{E}_T$ can be 
better observed in the tail of the distribution of the transverse mass of the 
lepton because, contrary to the SM background, the cross section of the signal 
grows with the energy due to the absence of propagator suppression.

A number of searches have been carried out at the LHC in this regard. Here, we 
focus on the ATLAS study of Ref.~\cite{Aaboud:2017efa}, based on $36$ fb$^{-1}$ 
of data collected at $\sqrt{s} = 13$~TeV. The main selection criteria are the 
requirement of exactly one light lepton with $p_T > 55$ GeV ($65$ GeV) in the 
muon (electron) channel; and likewise for $\slashed{E}_T$. The events are 
subsequently categorised according to the variable $m_T$, with $m_T^2 = 2p_T 
\slashed{E}_T (1-\cos{\phi_{\ell \slashed{E}_T}})$.

The numbers of predicted SM events and of observed 
events in four $m_T$ regions as provided in Ref.~\cite{Aaboud:2017efa} are 
shown in Tab.~\ref{tab:lepsLHC}. For convenience, we also show the maximum 
number of signal events, $s_\text{max}$, in each bin separately. We have 
obtained them using the CL$_s$ method~\cite{Read:2002hq} including the quoted 
uncertainties.

Following Eq.~\ref{eq:pplv}, the number of events in the signal side in any 
of the bins can be expressed as:
\begin{align}\label{eq:obs1}\nonumber
 N = \frac{1}{\Lambda^4}\bigg\lbrace& \left[(\alpha_{quNl}^{11i})^2+4 
(\alpha_{duNe}^{11i})^2+(\alpha_{lNqd}^{i11})^2\right]\mathcal{A}_1 + 
\left[4(\alpha_{duNe}^{11i})^2 + 
(\alpha_{ldqN}^{i11})^2\right]\mathcal{A}_2\\
&+2\left[4(\alpha_{duNe}^{11i})^2 - 
\alpha_{lNqd}^{i11}\,\alpha_{ldqN}^{i11}\right]\mathcal{A}_3 + 
(\alpha_{HNe}^i)^2\mathcal{A}_4 + (\alpha_{NW}^i)^2\mathcal{A}_5
\bigg\rbrace\,,
\end{align}
%
where $\mathcal{A}_1, \mathcal{A}_2, \cdots, \mathcal{A}_5$ are bin-dependent 
coefficients to be determined by 
simulation after recasting the 
experimental analysis. Our findings are also reported in 
Tab.~\ref{tab:lepsLHC}. The values of the $\mathcal{A}$ coefficients for 
electrons and muons differ only by a factor of $0.6$, introduced to simulate 
the smaller muon detection 
efficiency due to the strict trigger and muon selection criteria of the 
experimental analysis.
\begin{table}[t]
\begin{center}
\begin{tabular}{|c|ccc|}
  \hline
  & $400-600$ & $600-1000$ & $1000-2000$~[GeV]\\
  \hline
  $\mathcal{A}_1$ & $7400$ (4400) & 12000 (7200) & 16000 (9600)  
\\
  $\mathcal{A}_2$ & 2100 (1300) & 3600 (2200) & 4700 (2800)  \\
  $\mathcal{A}_3$ & $-3500$ ($-2100$) & $-5600$ ($-3400$) & $-7700$ ($-4600$)  
 \\  
  $\mathcal{A}_4$ & $5.0$ ($3.0$) & $1.2$ ($0.73$) & $0.15$ ($0.089$)  \\
  $\mathcal{A}_5$ & $360$ ($210$) & $210$ ($120$) & $79$ ($47$) \\
  \hline
  SM & $9700\pm 500 \; (6460 \pm 330)$ &$2010\pm 140 \; (1320 \pm 90)$ & $232\pm 
24 \; (150 \pm 13)$  \\
  data & 9551 (6772) &1931 (1392) & 246 (177) \\
  $s_\text{max}$ & 791 (778) & 213 (257) & 67 (62) \\
  \hline
 \end{tabular}
 \end{center}
 \caption{Coefficients in TeV$^4$ and rounded to two significant figures for 
$pp\to \ell N$ obtained upon recasting the experimental analysis of 
Ref.~\cite{Aaboud:2017efa} for $\mathcal{L} = 36$ fb$^{-1}$. The numbers outside (inside)
the parentheses refer to the $\ell = e \; (\mu)$ case.}
\label{tab:lepsLHC}
\end{table}
%
%
\begin{table}[ht!]
 \vspace{0.7cm}
 \begin{center}
 \begin{tabular}{|c|ccc|}
  \hline
  & $0-500$ & $500-1000$ & $>1000$~[GeV]\\
  \hline
  $\mathcal{B}_1$ & 170 & 3600 & 10000  \\
  $\mathcal{B}_2$ & 40 & 990 & 3200  \\
  $\mathcal{B}_3$ & $-69$ & $-1600$ & $-4700$  \\
  $\mathcal{B}_4$ & 0.33 & 1.0 & 0.15 \\
  $\mathcal{B}_5$ & 40 & 290  & 160\\ \hline
  SM & $1243\pm 160$ & $485\pm 77$ & $23.4\pm 6.2$ \\
  data & 1203 & $452$ & 15 \\
  $s_\text{max}$ & 258 & 125 & 12 \\
  \hline
 \end{tabular}
 \end{center}
 \caption{Coefficients in TeV$^{4}$ and rounded to two significant figures
for $pp\to\tau N$ obtained upon recasting the experimental analysis 
of Ref.~\cite{Sirunyan:2018lbg} for $\mathcal{L} = 35.9$ 
fb$^{-1}$. Note that Eq.~\ref{eq:pplv} is obtained 
under the assumption $\sqrt{s} \gg m_W$, so $\mathcal{B}_{4,5}$ in 
the first bin should not be taken rigorously.}
\label{tab:tausLHC}
\end{table}
%
For the same operators with taus instead of light leptons, we recast the CMS
analysis of Ref.~\cite{Sirunyan:2018lbg}, based on $35.9$ fb$^{-1}$ of 
data collected at $\sqrt{s} = 13$ TeV. More importantly, it requires a hadronic 
tau with $p_T>80$ GeV, as well as $\slashed{E}_T > 200$ GeV. This analysis 
divides the transverse mass range into slightly different bins; see 
Tab.~\ref{tab:tausLHC} which, as in the previous case, also reports the number 
of predicted SM events, the observed number of events and $s_\text{max}$, as 
well as the values of the different coefficients (this time dubbed 
$\mathcal{B}$) as obtained from simulation.

In principle, the same operators but involving second 
generation quarks also contribute (without interfering with the previous ones) 
to this process. However, the corresponding $\mathcal{A}$ and $\mathcal{B}$ 
coefficients are expected to be a factor 
of $\sim 10$ smaller due to the smaller parton distribution functions (PDFs).

\subsection{Monojet searches at the LHC}
\label{sec:monojet}
%
Four-fermion operators containing two light quarks and two RH 
neutrinos, \textit{i.e.} ${\cal O}_{uN}$, ${\cal O}_{dN}$, ${\cal O}_{qN}$, can 
lead to monojet searches at the LHC if for example a gluon is emitted from one 
of the initial quarks in $pp$ collisions.
They also modify the $\pi^0$ width but chirality 
suppressed. Furthermore, this width is experimentally known to be sensibly 
large~\cite{Tanabashi:2018oca}.

Different monojet searches have been performed by ATLAS and CMS. 
In this work, we consider the CMS analysis of Ref.~\cite{Sirunyan:2017jix}, 
based on $35.9$~fb$^{-1}$ of data collected at $\sqrt{s} = 13$~TeV. 
The main cuts of this study are missing transverse energy above $250$ GeV, at least one hard jet with $p_T> 250$ GeV and no isolated leptons. 
The analysis defines different signal 
regions depending on the value of $\slashed{E}_T$; see Tab.~\ref{tab:monojet}.

The interference between the three aforementioned operators is chirality 
suppressed. Therefore, in very good approximation, one can estimate the number 
of signal events in any region of the analysis as
\begin{equation}\label{eq:obs2}
 N = \frac{1}{\Lambda^4}\left[(\alpha_{uN}^{11})^2 \mathcal{C}_1 +  
(\alpha_{dN}^{11})^2 \mathcal{C}_2 +  
(\alpha_{qN}^{11})^2 \mathcal{C}_3\right]~,
\end{equation}
%
with $\mathcal{C}_1, \mathcal{C}_2, \mathcal{C}_3$ 
depending again on the signal region. 

Note that in this case we are neglecting the contribution of the operator 
${\cal O}_{HN}$, which in principle interferes with the previous ones, 
because it does not grow at large energies.
At any rate, as commented previously, we focus on scenarios with only 
four-fermion operators when setting bounds in section~\ref{sec:fit}.
Likewise, other four-fermion operators contribute to the monojet channel 
via $\nu N$. They of course do not interfere with the ones in Eq.~\ref{eq:obs2}. 
Besides, they are constrained by other observables, 
so we conservatively neglect them. 
\begin{table}[t]
 \begin{center}
 \begin{tabular}{|c|ccccc|}
  \hline
  & $690-740$ & $740-790$ & $790-840$ & $840-900$ & $900-960$~[GeV]\\
  \hline
  $\mathcal{C}_1$ & 210 & 170 & 130 & 130 & 94 \\
  $\mathcal{C}_2$ & 97 & 78 & 59 & 53& 39\\
  $\mathcal{C}_3$ & 320 & 250 & 180 & 170 & 130\\ \hline
  SM & $526\pm 14$ & $325\pm 12$ & $223\pm 9$ & $169\pm 8$ & $107\pm 6$ \\
  data & $557$ & $316$ & $233$ & $172$ & $101$ \\
  $s_\text{max}$ & 82 & 40 & 44 & 35 & 21\\
  \hline
 \end{tabular}
 \end{center}
 \caption{Coefficients in TeV$^4$ and rounded to two significant figures for 
$pp\rightarrow N\overline{N}g(q)$ obtained upon recasting the experimental 
analysis of Ref.~\cite{Sirunyan:2017jix} for $\mathcal{L} = 35.9$ fb$^{-1}$.}
\label{tab:monojet}
\end{table}

\subsection{Pion decays}
\label{sec:pion}
%
The operators modifying the tail of $\ell + \slashed{E}_T$ in $pp$ collisions 
can also enhance the pion decays, provided $N$ is 
the actual RH component of a Dirac neutrino.
(Also if $N$ is Majorana with low enough mass.)
For the computation of the corresponding matrix elements, 
we neglect $\mathcal{O}_{ldqN}^{i11}$, whose tensor form factor 
is hard to estimate. We obtain
\begin{align}\nonumber
 \mathcal{M}(\pi^-\rightarrow \ell_i^- \overline{N}) &= 
\frac{1}{\Lambda^2}
\overline{u(p_{\ell_i})}\bigg\lbrace
\langle 0|V^\mu|\pi^-\rangle 
\left[\left(\alpha_{duNe}^{11i} + \alpha_{HNe}^i\right)\gamma_\mu 
+ 2\sqrt{2}\, \frac{\alpha_{NW}^i}{m_W}\left(\gamma_\mu\slashed{p}-p_\mu\right)\right]\\
 &\phantom{{}={}\frac{1}{\Lambda^2}\overline{u(p_{\ell_i})}\bigg\lbrace}
+ \langle 0|S|\pi^-\rangle
\left(\alpha_{quNl}^{11i}-\alpha_{lNqd}^{i11}\right)
\bigg\rbrace P_R v(p_N)\,,
\end{align}
%
where $p = p_{\ell_i} + p_N$.
A similar expression holds for $\pi^+ \rightarrow \ell_i^+N$.

Following Ref.~\cite{Carpentier:2010ue}, we assume the expectation values of 
the quark currents to be determined by current algebra: $\langle 
0|V^\mu|\pi^\pm\rangle = f_\pi p^\mu$ and $\langle 0|S|\pi^\pm\rangle = f_\pi 
m_\pi^2/(m_u+m_d)$ with $f_\pi\sim 131$ MeV.
Then, the corresponding decay width reads
\begin{equation}\label{eq:obs3}
\Gamma(\pi^- \to \ell_i^- \overline{N}) = \frac{f_\pi^2 m_\pi}{16\pi\Lambda^4} 
\left(1-\frac{m_{\ell_i}^2}{m_\pi^2}\right)^2
\left[m_{\ell_i} \left(\alpha_{duNe}^{11i} + \alpha_{HNe}^i\right) 
+ \frac{m_\pi^2}{m_u+m_d} \left(\alpha_{quNl}^{11i} -\alpha_{lNqd}^{i11}\right)\right]^2.
\end{equation}
%

The measured values of the pion decay widths into electrons and muons read 
$\Gamma(\pi \rightarrow e + \text{inv}) = (310 \pm 1)\times10^{-23}$~GeV 
and $\Gamma(\pi \rightarrow \mu + \text{inv}) = (25279 \pm 
5)\times10^{-21}$~GeV, respectively~\cite{Tanabashi:2018oca}. 
We bound the coefficients entering this equation by requiring that the 
corresponding contribution is smaller than twice the experimental error.

\subsection{Tau decays}
\label{sec:tau}
%
The following operators give contributions to $\tau \rightarrow \ell + \mathrm{inv}$:
\begin{align}
%
	\mathcal{O}_{eN}^{i3} &= (\overline{\ell_{iR}} \gamma_\mu \tau_{R}) (\overline{N}\gamma^\mu N)\,,\\
	\mathcal{O}_{lN}^{i3} &= (\overline{\nu_{iL}}\gamma_\mu \nu_{\tau L}) (\overline{N}\gamma^\mu N) 
	+ \boldsymbol{(\overline{\ell_{iL}}\gamma_\mu \tau_{L}) (\overline{N}\gamma^\mu N)}\,,\\
	\mathcal{O}_{lNle}^{ii3} &= \boldsymbol{(\overline{\nu_{iL}} N)(\overline{\ell_{iL}} \tau_R)
	- (\overline{\ell_{iL}} N)(\overline{\nu_{iL}} \tau_R)} + \mathrm{h.c.}\,,\\
	\mathcal{O}_{lNle}^{i3i} &= \boldsymbol{(\overline{\nu_{iL}} N)(\overline{\tau_L} \ell_{iR})} 
	- (\overline{\ell_{iL}} N)(\overline{\nu_{\tau L}} \ell_{iR}) + \mathrm{h.c.}\,,\\
	\mathcal{O}_{lNle}^{3ii} &= (\overline{\nu_{\tau L}} N)(\overline{\ell_{iL}} \ell_{iR}) 
	-\boldsymbol{(\overline{\tau_L} N)(\overline{\nu_{iL}} \ell_{iR})} + \mathrm{h.c.}\,,\\
  	\mathcal{O}_{lNle}^{i33} &= (\overline{\nu_{iL}} N)(\overline{\tau_{L}} \tau_{R}) 
	-\boldsymbol{(\overline{\ell_{iL}} N)(\overline{\nu_{\tau L}} \tau_{R})} + \mathrm{h.c.}\,,\\
	\mathcal{O}_{lNle}^{3i3} &= \boldsymbol{(\overline{\nu_{\tau L}} N)(\overline{\ell_{iL}} \tau_{R})} 
	- (\overline{\tau_{L}} N)(\overline{\nu_{iL}} \tau_{R}) + \mathrm{h.c.}\,, \\
	\mathcal{O}_{lNle}^{33i} &= \boldsymbol{(\overline{\nu_{\tau L}} N)(\overline{\tau_{L}} \ell_{iR}) 
	- (\overline{\tau_{L}} N)(\overline{\nu_{\tau L}} \ell_{iR})} + \mathrm{h.c.}\,,\\
	\mathcal{O}_{HNe}^{i} &\supset \frac{v m_W}{\sqrt{2}} (\overline{N} \gamma_\mu \ell_{iR}) W^{+\mu} + \mathrm{h.c.}\,, \\
	\mathcal{O}_{HNe}^{3} &\supset \frac{v m_W}{\sqrt{2}} (\overline{N} \gamma_\mu \tau_{R}) W^{+\mu} + \mathrm{h.c.}\,, \\
	\mathcal{O}_{NW}^{i} &\supset 2v (\overline{\ell_{iL}} \sigma_{\mu\nu} N) \partial^\mu W^{-\nu} + \mathrm{h.c.}\,, \\
	\mathcal{O}_{NW}^{3} &\supset 2v (\overline{\tau_{L}} \sigma_{\mu\nu} N) \partial^\mu W^{-\nu} + \mathrm{h.c.}
\end{align}
%
When there are more than one term, only those in boldface are relevant. 
These operators lead to the following decay widths of the tau lepton:
\begin{align}
\label{eq:obs4}
\Gamma(\tau \rightarrow \ell_i \overline{N} N) &= \frac{m_\tau^5}{1536\pi^3\Lambda^4}
\left[(\alpha_{eN}^{i3})^2 + (\alpha_{lN}^{i3})^2\right], \\
\Gamma(\tau \rightarrow \ell_i \overline{N} \nu_i) &= \frac{m_\tau^5}{6144\pi^3\Lambda^4}
(\alpha_{lNle}^{ii3})^2\,, \\
\Gamma(\tau \rightarrow \ell_i \overline{\nu_i} N) &= \frac{m_\tau^5}{6144\pi^3\Lambda^4}
\bigg[(\alpha_{lNle}^{i3i})^2 + (\alpha_{lNle}^{3ii})^2 - \alpha_{lNle}^{i3i} \alpha_{lNle}^{3ii} 
+ 4 (\alpha_{HNe}^3)^2 \nonumber \\
&\phantom{{}={}\frac{m_\tau^5}{6144\pi^3\Lambda^4}\bigg[} 
+ \frac{64}{5}\frac{m_\tau^2}{m_W^2}(\alpha_{NW}^3)^2 
+ 8\sqrt{2}\frac{m_\tau}{m_W} \alpha_{HNe}^3 \alpha_{NW}^3\bigg]\,, \\
\Gamma(\tau \rightarrow \ell_i \overline{N} \nu_\tau) &= \frac{m_\tau^5}{6144\pi^3\Lambda^4}
\bigg[(\alpha_{lNle}^{i33})^2 + (\alpha_{lNle}^{3i3})^2 - \alpha_{lNle}^{i33}\,\alpha_{lNle}^{3i3} 
+ 4 (\alpha_{HNe}^i)^2 \nonumber \\
&\phantom{{}={}\frac{m_\tau^5}{6144\pi^3\Lambda^4}\bigg[} 
+ \frac{24}{5}\frac{m_\tau^2}{m_W^2}(\alpha_{NW}^i)^2 
- 2\sqrt{2}\frac{m_\tau}{m_W} \alpha_{lNle}^{3i3} \alpha_{NW}^i\bigg]\,,\\
\Gamma(\tau \rightarrow \ell_i \overline{\nu_\tau} N) &= \frac{m_\tau^5}{6144\pi^3\Lambda^4}
(\alpha_{lNle}^{33i})^2\,.
\label{eq:obs4bis}
\end{align}
%

The measured values of the tau decay widths into electrons and muons are $\Gamma(\tau \to e + \text{inv}) = (4.03\pm 0.02)\times 10^{-13}$ GeV and $\Gamma(\tau \to \mu+\text{inv}) = (3.93\pm 0.02)\times 
10^{-13}$ GeV, respectively~\cite{Tanabashi:2018oca}. Following the same procedure as in the pion case, we bound these operators by requiring that the corresponding theoretical decay width is not larger than twice the experimental error.

\subsection{Top decays}
%
Finally, the following four-fermion operators contribute to flavour-conserving top decays:
\begin{align}\label{eq:topop}
\mathcal{O}_{duNe}^{33i} &= (\overline{b_R} \gamma_\mu t_R)
(\overline{N} \gamma^\mu \ell_{iR}) + \mathrm{h.c.}\,,\\
\mathcal{O}_{ldqN}^{i33} &= \boldsymbol{\frac{1}{2}(\overline{t_L} b_R) (\overline{\ell_{iL}} N)}
- \frac{1}{2}(\overline{b_L} b_R) (\overline{\nu_{iL}} N) \nonumber\\
&\phantom{{}={}} + \boldsymbol{\frac{1}{8}(\overline{t_L} \sigma_{\mu\nu} b_R) (\overline{\ell_{iL}} \sigma^{\mu\nu} N)} 
- \frac{1}{8}(\overline{b_L} \sigma_{\mu\nu} b_R) (\overline{\nu_{iL}} \sigma^{\mu\nu} N)
+\mathrm{h.c.}\,,\\
\mathcal{O}_{lNqd}^{i33} &= (\overline{\nu_{iL}} N) (\overline{b_L} b_R) 
- \boldsymbol{(\overline{\ell_{iL}} N) (\overline{t_L} b_R)} +\mathrm{h.c.}\,,\\
\mathcal{O}_{quNl}^{33i} &= (\overline{t_L} t_R) (\overline{N} \nu_{iL}) 
+ \boldsymbol{(\overline{b_L} t_R) (\overline{N} \ell_{iL})} + \mathrm{h.c.}
\end{align}
%
When there are more than one term, only those in boldface are relevant. 
The corresponding decay width reads
\begin{align}
\Gamma(t \rightarrow b \ell_i^+ N) &= \frac{m_t^5}{6144 \pi^3 \Lambda^4} 
\bigg[4(\alpha_{duNe}^{33i})^2 + (\alpha_{quNl}^{33i})^2 \nonumber \\
&\phantom{{}={}\frac{m_t^5}{6144\pi^3\Lambda^4}\bigg[} 
+ (\alpha_{ldqN}^{i33})^2 + (\alpha_{lNqd}^{i33})^2 - \alpha_{ldqN}^{i33} \alpha_{lNqd}^{i33}
\bigg]\,,
\label{eq:tTOblN}
\end{align}
%
and similarly for $\overline{t} \rightarrow \overline{b} \ell_i^- \overline{N}$.
In this expression, we have neglected the contribution of $\mathcal{O}_{HNe}$ 
and $\mathcal{O}_{NW}$. The reason is that in those cases the top decay is 
approximately two body,
followed by the leptonic decay of an on-shell $W$. This does not only allow to 
disentangle the two contributions, but the interference is also very small. 
Similar results have been pointed out in the SMEFT; see \textit{e.g.} 
Ref.~\cite{AguilarSaavedra:2010zi}.

These operators can be mostly probed only in top decays. In particular, $\mathcal{O}_{duNe}$ does not provide any other interaction. 
This is in contrast with analogous operators in the SMEFT, such as for example 
$\sim (\overline{t_R}\gamma_\mu t_R) (\overline{l_L}\gamma^\mu l_L)$, 
because the left-handed (LH) neutrinos always come along with charged leptons. 
These operators are therefore better tested 
at lepton facilities; see \textit{e.g.} Ref.~\cite{Durieux:2018tev}.
 
This discussion applies also to the flavour-violating top operators:
\begin{align}
\mathcal{O}_{uN}^{13} &= \boldsymbol{(\overline{u_R} \gamma_\mu t_R)(\overline{N}\gamma^\mu N)}\,, \\
\mathcal{O}_{qN}^{13} &= \boldsymbol{(\overline{u_L} \gamma_\mu t_L)(\overline{N}\gamma^\mu N)} + (\overline{d_L} \gamma_\mu b_L)(\overline{N}\gamma^\mu N)\,, \\
\mathcal{O}_{quNl}^{13i} &= \boldsymbol{(\overline{u_L} t_R)(\overline{N} \nu_{iL})} 
+ (\overline{d_L} t_R)(\overline{N} \ell_{iL}) + \mathrm{h.c.}\,, \\
 \mathcal{O}_{quNl}^{31i} &= \boldsymbol{(\overline{t_L} u_R)(\overline{N} \nu_{iL})} 
+ (\overline{b_L} u_R)(\overline{N} \ell_{iL}) + \mathrm{h.c.}
\end{align}
%
The terms in boldface lead to the rare top decays $t\rightarrow j +\text{inv}$ (where $j$ stands for $jet$); again without charged
lepton counterpart unlike the SMEFT~\cite{Durieux:2014xla,Chala:2018agk}. 
The corresponding decay widths read:
\begin{align}
\Gamma(t \rightarrow u \overline{N} N) &= \frac{m_t^5}{1536\pi^3\Lambda^4}
\left[(\alpha_{uN}^{13})^2 + (\alpha_{qN}^{13})^2\right], \\
\Gamma(t \rightarrow u \overline{\nu_i} N) &= \frac{m_t^5}{6144\pi^3\Lambda^4}
(\alpha_{quNl}^{13i})^2\,, \\
\Gamma(t \rightarrow u \overline{N} \nu_i) &= \frac{m_t^5}{6144\pi^3\Lambda^4}
(\alpha_{quNl}^{31i})^2\,.
\end{align}
%

Similar expressions hold of course for second generation quarks. The sensitivity of measurements at colliders is expected to be similar in both cases, just as in other flavour-violating top decays~\cite{Papaefstathiou:2017xuv,Banerjee:2018fsx,Chala:2018agk,Chiang:2018oyd,
Jain:2019ebq,Altmannshofer:2019ogm}.

Current bounds on the top width are not constraining enough 
in any of these cases, though. 
Instead, dedicated analyses at colliders are to be performed 
to bound these operators. 
In section~\ref{sec:lhc} we develop one such analysis for the flavour-conserving 
top decay at the LHC. The flavour-violating processes 
involve a light quark and two sources of missing energy, 
making these much appropriate search channels at future lepton colliders, 
in which all the three components of the missing momentum are measured.

\section{Global constraints}
\label{sec:fit}
%
Taking into account the observables computed in Eqs.~\ref{eq:obs1} (for light 
leptons and taus; see Tabs.~\ref{tab:lepsLHC} and \ref{tab:tausLHC}, 
respectively), \ref{eq:obs2}, \ref{eq:obs3} and \ref{eq:obs4}--\ref{eq:obs4bis}, 
we can set 
constraints on the different 
operators of Tab.~\ref{tab:basis}.
The bounds are given by the non-boldfaced numbers in 
Tabs.~\ref{tab:bounds1}--\ref{tab:bounds3}.

When relying on high energy searches at colliders, we  
derive the bounds on the operator coefficients from the several bins for each 
of the analyses and keep the bounds yielding the 
strongest result. However, we also ensure that we do not go very high in 
energies in order to abide by EFT validities.
Specifically, we remain within energy bins of less than 1 TeV. 

Deriving the bounds on 
$\alpha_{uN}^{11}, \alpha_{dN}^{11}$ and $\alpha_{qN}^{11}$ is particularly 
simple as they enter the analytical expression for the number of events, for 
the monojet analysis, without any interference. In this case, the 
$\slashed{E}_T$ bin of $[740-790]$ GeV yields the strongest limits. 

For the 
$\ell + \slashed{E}_T$ analysis, it is straightforward to constrain 
$\alpha_{quNl}^{11i}$ and $\alpha_{duNe}^{11i}$ as they do not interfere with
any operator 
in the expression for the number of events. For the 
operators that do interfere, \textit{viz.} 
$\mathcal{O}_{lNqd}^{i11}$ and $\mathcal{O}_{ldqN}^{i11}$, we marginalise over 
each coefficient to set a bound on the second one. Thus, in order to constrain 
$\alpha_{ldqN}^{i11}$, we fix $\alpha_{lNqd}^{i11}$ to the value that 
minimises the number of events for each value of $\alpha_{ldqN}^{i11}$, from 
where the procedure of setting bounds reduces to the one dimension as in the 
previous cases.
The reverse process can be followed to constrain $\alpha_{lNqd}^{i11}$.

The bounds obtained this way must be 
taken if $N$ is a Majorana neutrino with mass above $m_\pi$. 
For Dirac neutrinos, we bound $\alpha_{lNqd}^{i11}$ instead by using the pion 
decay width in Eq.~\ref{eq:obs3} (neglecting the $m_{\ell_i}$ piece). The 
corresponding limit is a factor of $1.9$ ($1.8$) more stringent for electrons 
(muons) than the 
LHC counterpart. This is the constraint we show in tables. Operators involving 
electrons are more constrained than those involving 
muons due to the larger 
sensitivity of the analysis of Ref.~\cite{Aaboud:2017efa} to electrons. 
Also, the pion decay width to $\mu+\text{inv}$ is significantly larger than that for 
$e+\text{inv}$, therefore leaving more space for new physics.

The $\tau + \slashed{E}_T$ follows a similar technique and helps us bound
$\alpha_{quNl}^{113}, \alpha_{duNe}^{113}, \alpha_{ldqN}^{311}$ and 
$\alpha_{lNqd}^{311}$. We use the respective bins of $[600-1000]$~GeV in 
$m_T$ for the $\ell + \slashed{E}_T$ analysis and $[500-1000]$~GeV 
for the $\tau + \slashed{E}_T$ case. 
When the aforementioned operators are considered with second generation 
quarks instead of the first family, the bounds get weaker 
by a factor of $\sim 3$. Thus, without relying on flavour observables, one can 
estimate the bounds on operators involving transitions between the first and 
second quark families to be those quoted in the tables up to a factor of $\sim 1-3$. 
(Flavour observables can be much more constraining, though.)

Finally, we bound operators modifying the tau decays into electrons and muons. 
$\mathcal{O}_{eN}^{i3}$, $\mathcal{O}_{lN}^{i3}$, $\mathcal{O}_{lNle}^{ii3}$ 
and $\mathcal{O}_{lNle}^{33i}$ can be straightforwardly constrained, because 
they do not interfere with any other operator. $\mathcal{O}_{lNle}^{i3i}$ and 
$\mathcal{O}_{lNle}^{3ii}$ interfere among themselves; we bound each by 
marginalising over the other. Likewise for $\mathcal{O}_{lNle}^{i33}$ and 
$\mathcal{O}_{lNle}^{3i3}$. In this case, muonic and electronic operators get 
more or less equally constrained. Similar operators but involving transitions 
between the first and second lepton families could be constrained using muon 
decays. Naively, the bounds would be a factor of $10 - 100$ stronger. 
However, the Fermi constant would be also redefined on account 
of the Wilson coefficients, making the process of bounding these operators 
more subtle.

Among those operators not constrained by the observables considered in this 
paper, we find $\mathcal{O}_{NN}$, which only involves RH neutrinos.
We also have $\mathcal{O}_{eN}^{ii}$ and $\mathcal{O}_{lN}^{ii}$; 
these could be tested in monophoton searches at lepton colliders. Third 
generation flavour conserving 
$\mathcal{O}_{dN}^{33}$, $\mathcal{O}_{uN}^{33}$ and $\mathcal{O}_{qN}^{33}$
could be tested in searches for $b\overline{b} + \slashed{E}_T$ and
$t \overline{t} + \slashed{E}_T$.
Most of the rest of operators induce new top decays, either into 
$b\ell+\text{inv}$, $b\tau+\text{inv}$ or $j+\text{inv}$. The latter one is 
hard to probe at hadron colliders, because it involves two sources of missing 
energy and only light jets. (Moreover, is flavour-violating.) The second one 
involves a tau lepton, making this signal less promising than the first one. 
The first signal gives rise to a final state which is identical to the SM leptonic 
top decay (which complicates its study), 
but interestingly in this case the lepton and the missing energy do 
not reconstruct a $W$ boson. This provides a completely new signal not yet 
explored experimentally, for which we design a completely novel search strategy. 
For completeness, we advance the prospective bounds on the corresponding 
operators by the boldfaced numbers in Tabs.~\ref{tab:bounds1} and 
\ref{tab:bounds3}.
\begin{table}[t]
 \begin{center}
 \begin{tabular}[t]{|l|ccc|}
 \hline
  Operator & $\alpha_\mathrm{max}$ for $\Lambda = 1$~TeV & $\Lambda_\mathrm{min}$~[TeV] for $\alpha = 1$ & Observable \\
  \hline
  ${\cal O}_{eN}^{i3}$ & 3.0~(2.9) & 0.58~(0.59) & $\tau \rightarrow \ell + \mathrm{inv}$ \\
  \hdashline
  ${\cal O}_{dN}^{11}$ & 0.72 & 1.2 & monojet \\
  \hdashline
  ${\cal O}_{uN}^{11}$ & 0.48 & 1.4 & monojet \\
  \hdashline
  ${\cal O}_{duNe}^{11i}$ & 0.11~(0.16) & 3.0~(2.5) & $\ell + \slashed{E}_T$ \\
  ${\cal O}_{duNe}^{113}$ & 0.15 & 2.6 & $\tau + \slashed{E}_T$ \\
  ${\cal O}_{duNe}^{33i}$ & \textbf{9.2 (9.2)} & \textbf{0.33 (0.33)} & $\boldsymbol{t \to b\ell +\mathrm{inv}}$\\
  \hline
 \end{tabular}
\caption{\it Maximum (minimum) value of $\alpha$ ($\Lambda$) for $\Lambda = 1$ 
TeV ($\alpha = 1$) allowed by the observables quoted in the 
last column for $RRRR$ operators. 
The numbers outside (inside) the parentheses refer to $\ell = e~(\mu)$.}
\label{tab:bounds1}
 \end{center}
\end{table}
%
%
\begin{table}[t]
 \begin{center}
 \begin{tabular}[t]{|l|ccc|}
 \hline
  Operator & $\alpha_\mathrm{max}$ for $\Lambda = 1$~TeV & 
  $\Lambda_\mathrm{min}$~[TeV] for $\alpha = 1$ & Observable \\
  \hline
  ${\cal O}_{lN}^{i3}$ & 3.0~(2.9) & 0.58~(0.59) & $\tau \rightarrow \ell + \mathrm{inv}$ \\
  \hdashline
  ${\cal O}_{qN}^{11}$ & 0.40 & 1.6 &  monojet  \\
  \hline
 \end{tabular}
\caption{\it Maximum (minimum) value of $\alpha$ ($\Lambda$) for $\Lambda = 1$ 
TeV ($\alpha = 1$) allowed by the observables quoted in the 
last column for $LLRR$ operators. 
The numbers outside (inside) the parentheses refer to $\ell = e~(\mu)$.}
\label{tab:bounds2}
\end{center}
\end{table}
%
%
\begin{table}[t!]
 \vspace{0.5cm}
 \begin{center}
 \begin{tabular}[t]{|l|ccc|}
 \hline
  Operator & $\alpha_\mathrm{max}$ for $\Lambda = 1$~TeV & 
  $\Lambda_\mathrm{min}$~[TeV] for $\alpha = 1$ & Observable \\
  \hline
  ${\cal O}_{lNle}^{ii3}$ & 6.0~(5.9) & 0.41~(0.41) & $\tau \rightarrow \ell + \mathrm{inv}$\\
  ${\cal O}_{lNle}^{i3i}$ & 6.8~(6.8) & 0.38~(0.38) & $\tau 
\rightarrow \ell + \mathrm{inv}$\\
  ${\cal O}_{lNle}^{i33}$ & 6.8~(6.8) & 0.38~(0.38) & $\tau 
\rightarrow \ell + \mathrm{inv}$\\
  ${\cal O}_{lNle}^{3ii}$ & 6.8~(6.8) & 0.38~(0.38) & $\tau 
\rightarrow \ell + \mathrm{inv}$\\
  ${\cal O}_{lNle}^{3i3}$ & 6.8~(6.8) & 0.38~(0.38) & $\tau 
\rightarrow \ell + \mathrm{inv}$\\
  ${\cal O}_{lNle}^{33i}$ & 6.0~(5.9) & 0.41~(0.41) & $\tau \rightarrow \ell + \mathrm{inv}$\\
  \hdashline
  ${\cal O}_{ldqN}^{i11}$ & 0.46~(0.66) & 1.5~(1.2) & $\ell + 
\slashed{E}_T$ \\
  ${\cal O}_{ldqN}^{i33}$ & \textbf{21 (21)} & \textbf{0.22 (0.22)} & $\boldsymbol{t \to b\ell + \mathrm{inv}}$ \\
  ${\cal O}_{ldqN}^{311}$ & 0.67 & 1.2 & $\tau + \slashed{E}_T$ \\
  \hdashline
  ${\cal O}_{lNqd}^{i11}$ & 0.25~(0.36) & 2.0~(1.7) & $\pi 
\rightarrow \ell + \mathrm{inv}$ \\
  ${\cal O}_{lNqd}^{i33}$ & \textbf{21 (21)} & \textbf{0.22 (0.22)} & $\boldsymbol{t \to b \ell +\mathrm{inv}}$ \\
  ${\cal O}_{lNqd}^{311}$ & 0.35 & 1.7 & $\tau + \slashed{E}_T$\\
  \hdashline
  ${\cal O}_{quNl}^{11i}$ & 0.13~(0.19) & 2.8~(2.3) & $\ell + \slashed{E}_T$ \\
  ${\cal O}_{quNl}^{113}$ & 0.19 & 2.3 & $\tau + \slashed{E}_T$\\
  ${\cal O}_{quNl}^{33i}$ & \textbf{18 (18)} & \textbf{0.23 (0.23)} & $\boldsymbol{t \to b \ell +\mathrm{inv}}$ \\
  \hline
 \end{tabular}
\caption{\it Maximum (minimum) value of $\alpha$ ($\Lambda$) for $\Lambda = 1$ 
TeV ($\alpha = 1$) allowed by the observables quoted in the 
last column for $LRRL$ operators. 
The numbers outside (inside) the parentheses refer to $\ell = e~(\mu)$.}
\label{tab:bounds3}
 \end{center}
\end{table}

\section{$t\rightarrow b\ell N$ at the LHC}
\label{sec:lhc}
%
We can constrain the operators $\mathcal{O}_{duNe}$, $\mathcal{O}_{\ell d qN}$, 
$\mathcal{O}_{\ell N q d}$ and $\mathcal{O}_{q u \ell N}$ in $t\bar{t}$ 
production with one of the tops decaying exactly as in the SM in the hadronic 
mode and the other decaying leptonically through the modified vertex. We 
focus on the high luminosity run of the LHC at 14 TeV with an integrated 
luminosity of 3 ab$^{-1}$. The final state consists of $b\bar{b}\ell 
\slashed{E}_T+$ jets. (We perform our analysis only in the muonic final state. 
We assume the results not to be significantly different for the case of 
electrons.)
The dominant background to this process comes from the SM semi-leptonic $t\bar{t}$ channel.
It is a very challenging task to reduce the background, as the topology is
essentially the same in both the signal and the background. In particular, the hadronic 
top decays in the exact same manner in both cases.

Our analysis strategy starts by generating the signal and background events at 
$\sqrt{s} = 14$~TeV using the NNPDF23\@LO PDF set\cite{Ball:2012cx}. We generate our 
samples at the leading order and then multiply by a flat next-to-next-to leading order (NNLO) + 
next-to-next-to leading logarithm (NNLL) $k$-factor of $\sim 1.63$ corresponding to 
$m_t = 172.5$ GeV and for the central value of the cross section. The central value of the 
14 TeV NNLO+NNLL cross section is 984.50 pb~\cite{Czakon:2011xx}. At the generation level, 
we demand $p_T(j,b,\ell) > 15, 15, 10$ GeV, $|\eta(j,b,\ell)| < 5,3,3$ 
and $\Delta R(jj,bb,bj,j\ell,b\ell) > 0.3$ and $\Delta R(\ell\ell) > 0.2$. We 
shower the events using \texttt{Pythia v8}~\cite{Sjostrand:2001yu, Sjostrand:2014zea}. 
We construct jets upon using the anti-k$_t$ algorithm~\cite{Cacciari:2008gp} with a jet 
parameter $R=0.4$ in the \texttt{FastJet v3} framework~\cite{Cacciari:2011ma}. We require 
the jets to have $p_T > 30$ GeV and the $b$-tagged jets to have $|\eta| < 2.5$. The leptons 
are required to have $p_T > 10$ GeV and $|\eta| < 2.5$. The $b$-tagged jets are 
constructed by requiring the $B$-meson tracks to be within $\Delta R = 0.2$ of a 
jet. Moreover, we implement a flat $b$-tagging efficiency of 70\%. In our 
simplified analysis, we also consider $c$ quarks and light quarks faking a 
$b$-jet with probabilities of 10\% and 1\%, respectively. Finally, isolated 
leptons are defined by requiring that the hadronic activity around $\Delta R = 0.2$ of the 
corresponding lepton is smaller than 10\% of its $p_T$.

Subsequently, we demand exactly two $b$-tagged jets, one isolated lepton and at 
least two light jets. 
In order to ascertain proper top-mass reconstruction, we build two hadronic 
top mass variables ($m_{had,1}^t$ and $m_{had,2}^t$). We include the two hardest 
light jets ($j_{1,2}$) when reconstructing both $m_{had,1}^t$ and 
$m_{had,2}^t$. However, we include the harder (softer) $b$-tagged jet to 
reconstruct $m_{had,1}^t$ ($m_{had,2}^t$). For each event we check which of 
these reconstructed hadronic top masses is closest to the actual top mass, 
that we take to be $m_t = 172.5$~GeV. The $b$-jet giving the poorer 
reconstruction is assigned to the leptonic top.
Next, we demand that the 
best reconstructed hadronic top ($m_t^{had}$) and the hadronic $W$ 
($m_W^{had}$), reconstructed out of the two hardest light jets, lie respectively 
within 40 GeV and 30 GeV of the top and $W$ masses (with $m_W$ taken as 80.385 GeV).

We reconstruct the $p_z$ of the neutrino by solving for the leptonic top mass.
There is a two-fold ambiguity in this process, which results into two values for 
the momentum of the $\ell \nu$ system, corresponding to the ``$+$'' and ``$-$'' 
solutions of the quadratic equation. We denote these two solutions by $W_1$ and 
$W_2$, respectively. (Note, however, that in the signal they do not need to be 
close to the $W$ mass.) 
We only select those events where the 
solution to the quadratic equation for the top mass has a positive discriminant. 
The aforementioned trigger cuts, isolation cuts and analysis cuts have the 
respective efficiencies of 4.4\% and 4.7\% for 
the signal and the SM background. 
\begin{figure}[t]
\centering
\includegraphics[width=.49\textwidth]{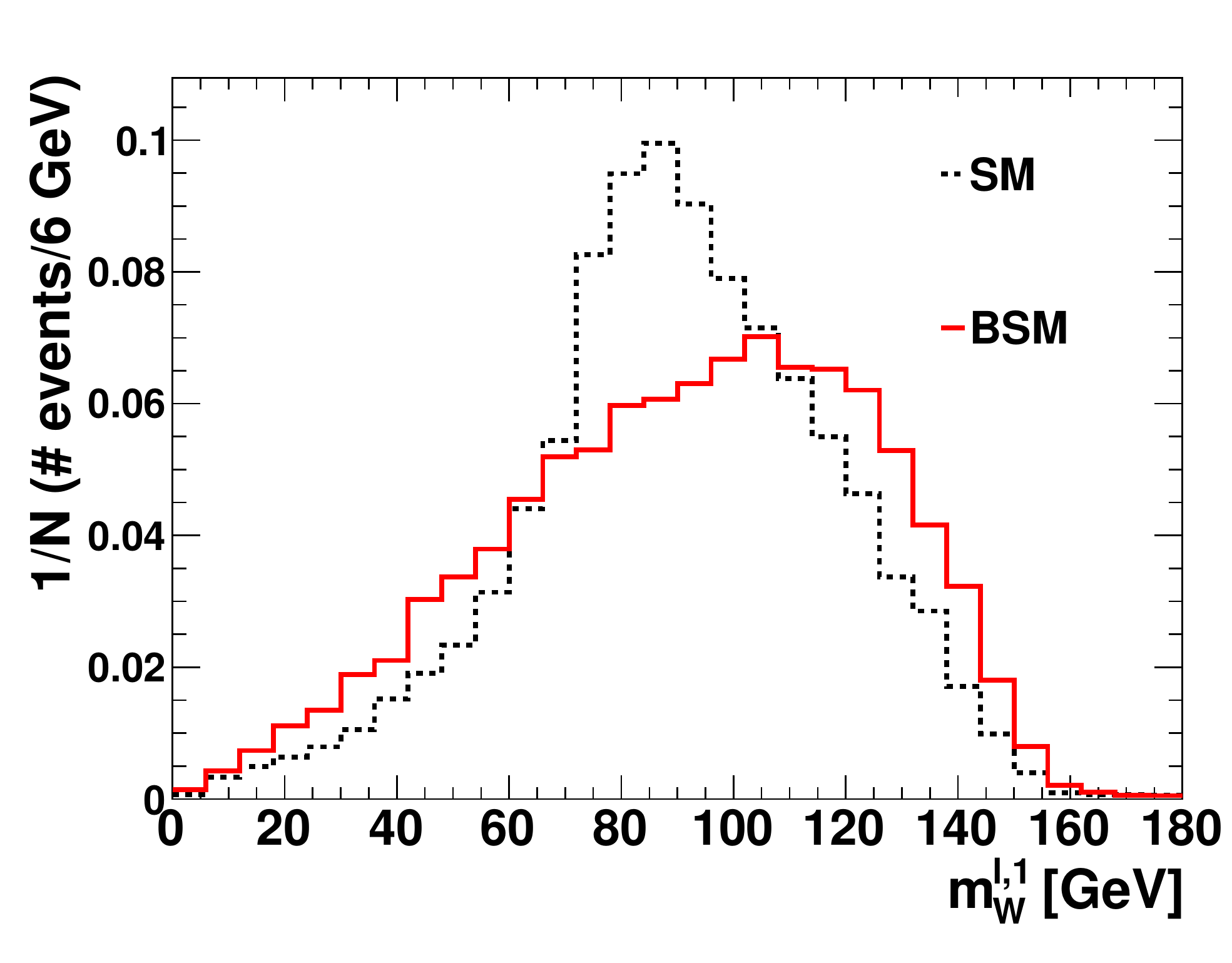}
\includegraphics[width=.49\textwidth]{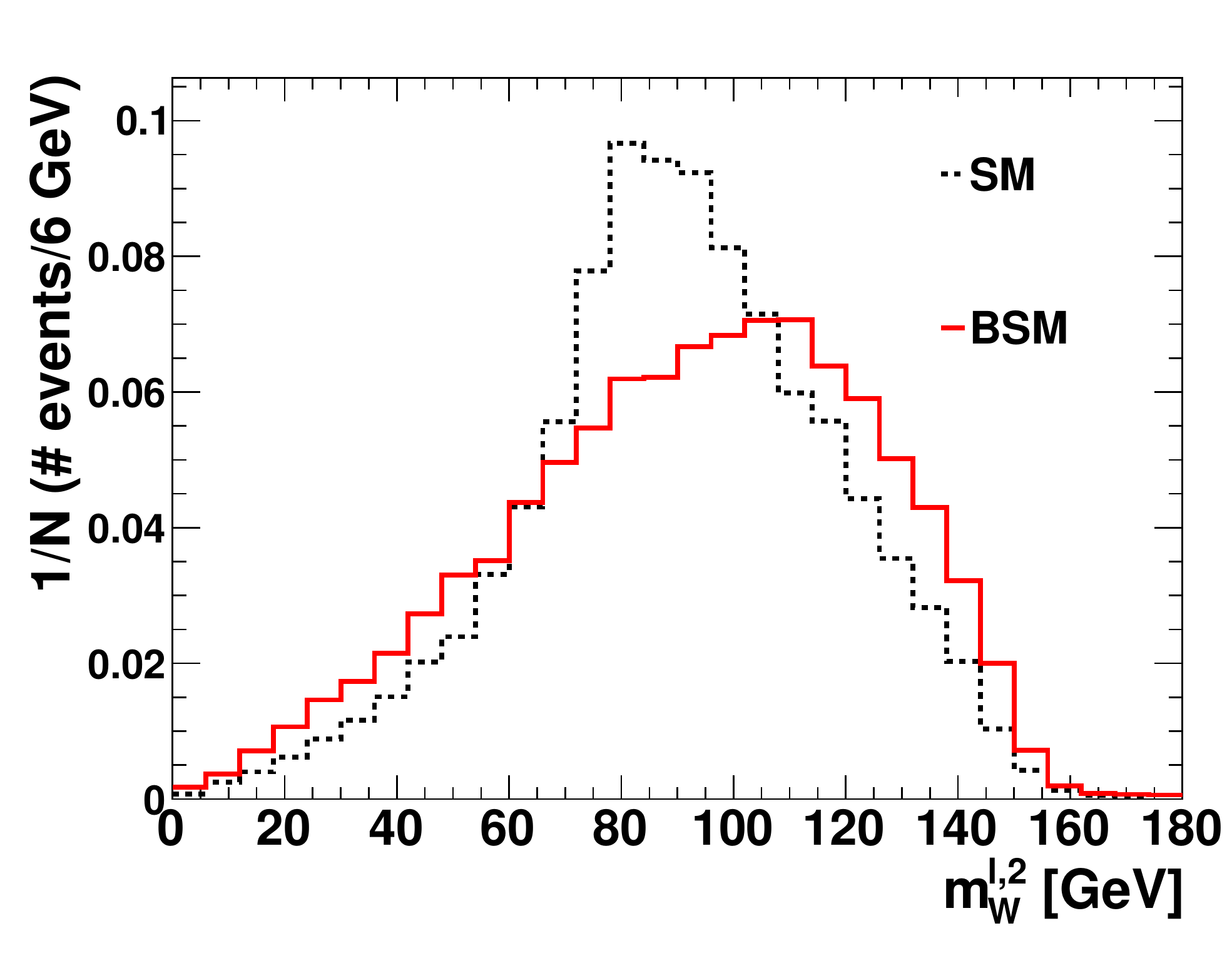}
\caption{\it 
Reconstructed $W$-boson mass for the signal (solid red) and background (dashed black).
Here 1 (2) refers to the ``+" (``$-$") solution of the
neutrino $p_z$ obtained while solving for the invariant
mass of the leptonic top.
}
\label{fig:LHCanalysis1}
\end{figure}
%
%
\begin{figure}[t!]
\vspace{1cm}
\centering
\includegraphics[width=.6\textwidth]{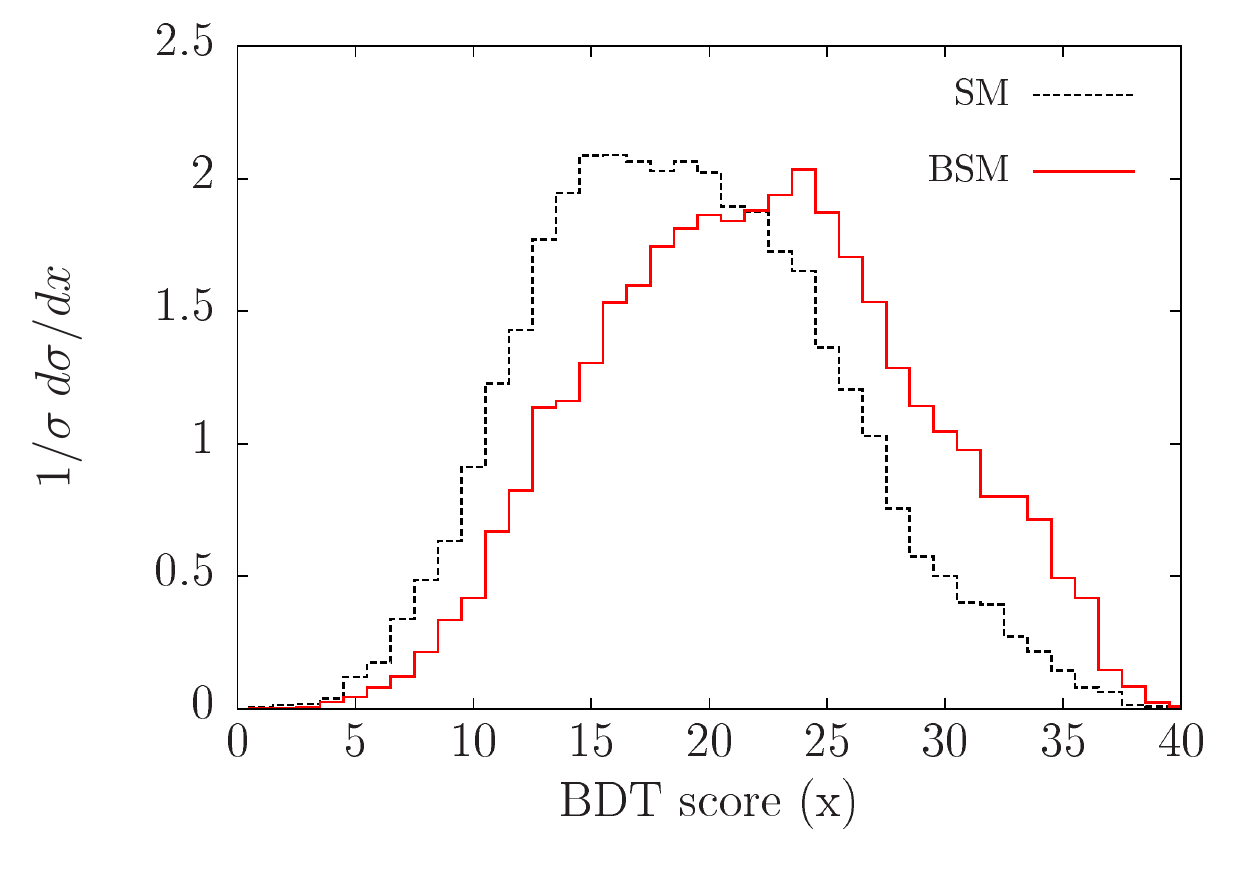}
\caption{\it 
BDT score variable used in the asymmetry determination 
as described in the text.}
\label{fig:LHCanalysis2}
\end{figure}
%
%
\begin{figure}[t]
 \includegraphics[width=0.5\columnwidth]{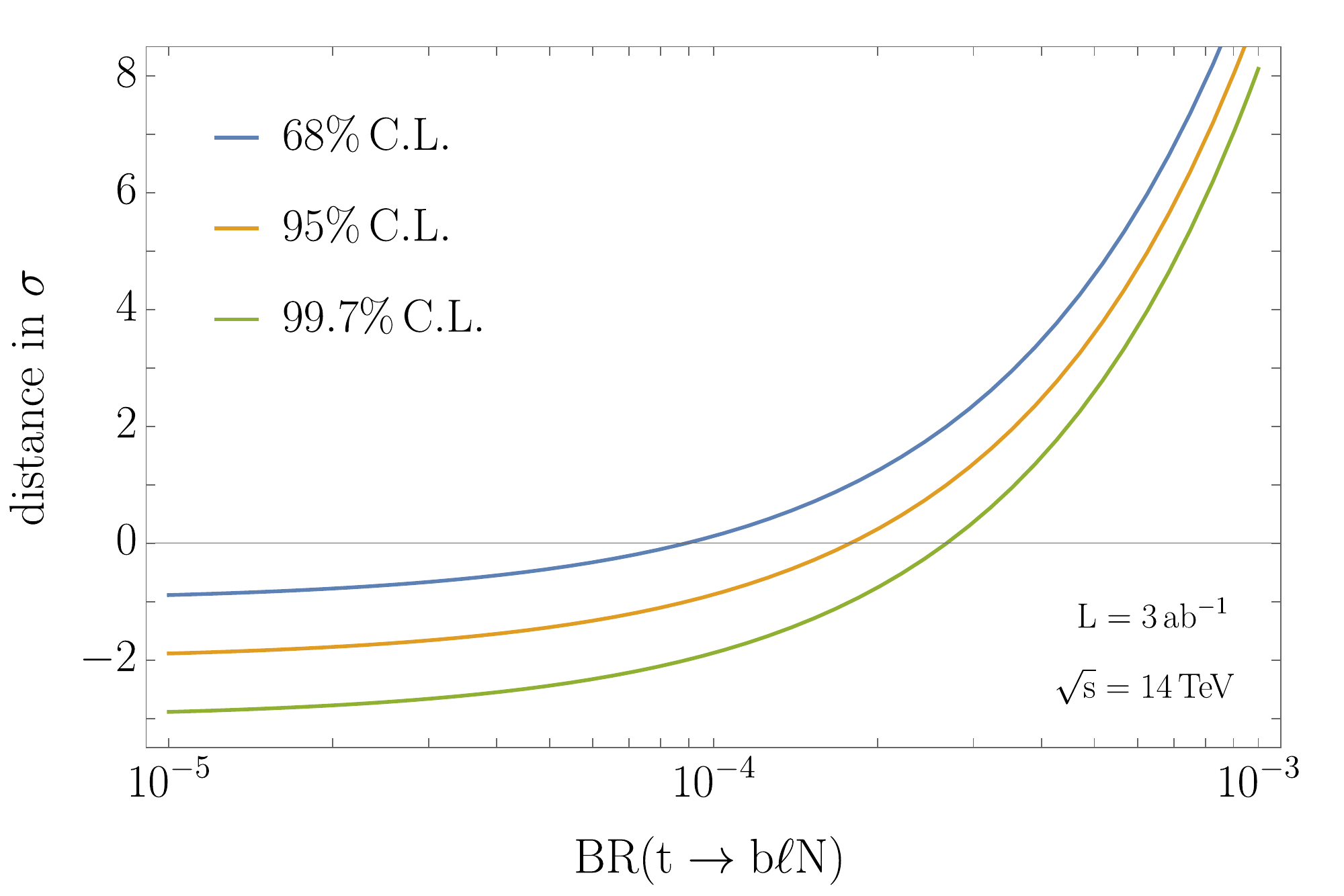}
 \includegraphics[width=0.5\columnwidth]{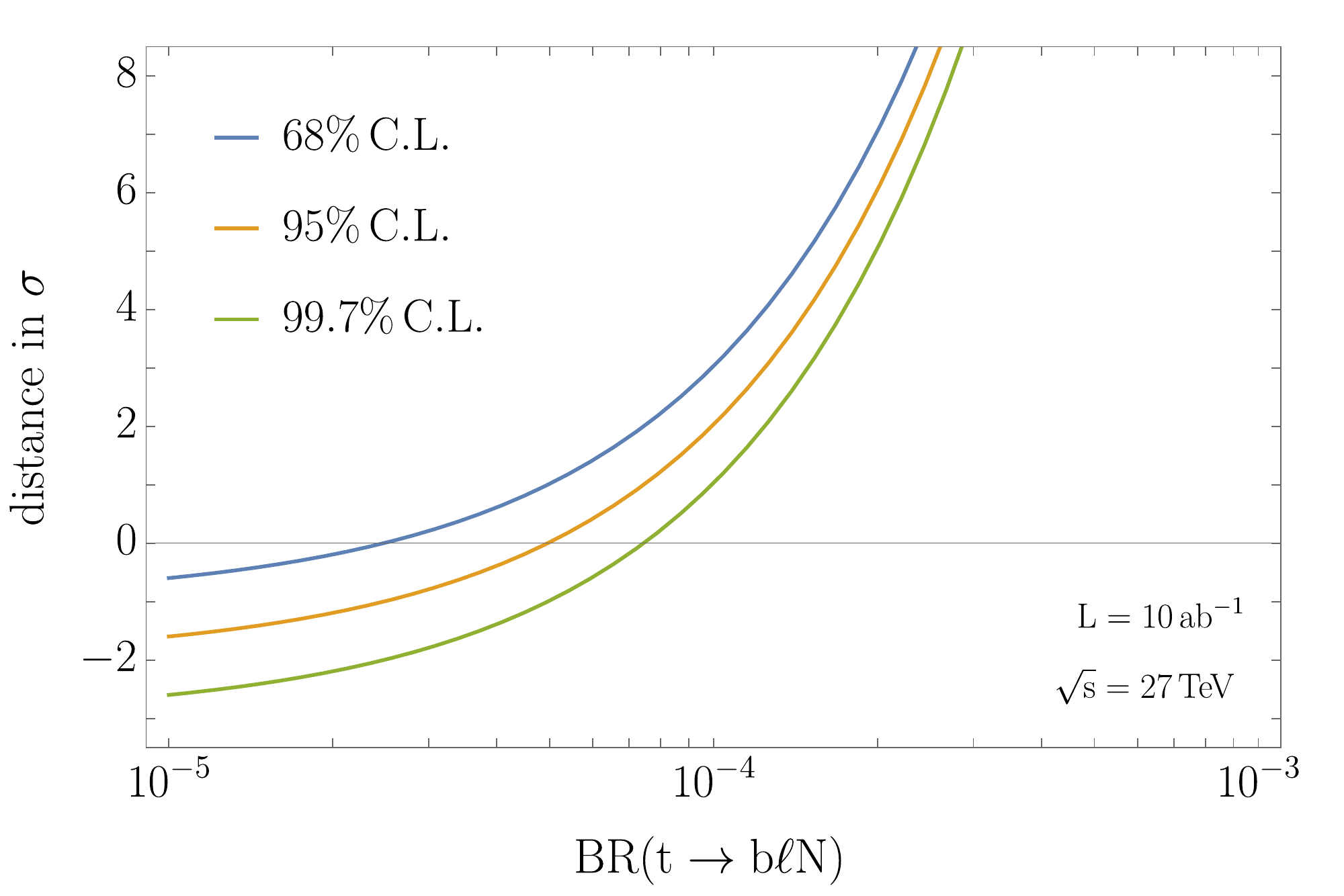}
 \caption{\it Number of standard deviations between the 
mean value of the BDT asymmetry in the signal + background 
and the asymmetry in the SM alone as a function of the exotic top decay mode
branching ratio for $\sqrt{s} = 14$ TeV and $\mathcal{L} = 
3$ ab$^{-1}$ (left) and for $\sqrt{s} = 27$ TeV and 
$\mathcal{L} = 10$ ab$^{-1}$ (right).}
\label{fig:finalTOP}
\end{figure}
%

The distribution of $m_{W}^{\ell,1/2}$, where 1 (2) refers to the 
``+" (``$-$") solution of the aforementioned quadratic equation,
is shown in Fig.~\ref{fig:LHCanalysis1}. Despite the difference in 
$m_{W}^{\ell,1/2}$ (that relies on the fact that there is no $W$ for the 
leptonic top in the EFT scenario), the signal and the background distributions 
are apparently not very different from the point of view of a cut-and-count 
analysis. We therefore perform a simple 
boosted decision tree (BDT)
analysis in the TMVA framework~\cite{Hocker:2007ht}. 
We include the following observables: the four-momenta of the lepton and the 
$b$-tagged jet best reconstructing the leptonic top, 
the two components of the transverse missing momentum, 
$m_W^{had}$, $m_{W}^{\ell,1}$, $m_{W}^{\ell,2}$, $m_t^{had}$, 
$\slashed{E}_T$, $\Delta R_{b, b}, \Delta R_{b_{\ell}, 
j_{1/2}}, \Delta R_{\ell, b_{h}}$ and $\Delta R_{\ell, j_{1/2}}$, where $b_{\ell} \; 
(b_{h})$ is the $b$-jet associated with the leptonic (hadronic) top. We 
ensure that there is no overtraining by checking that the Kolmogorov-Smirnov 
statistic for neither the signal nor the 
background falls below the critical value of $0.01$~\cite{KS}.
As expected, $m_{W}^{\ell,1}$ and $m_{W}^{\ell,2}$ serve as the best discriminating variables. The other variables
do not show significant prowess in the discriminating procedure.

We get a statistical significance of $\sim 13$ for BR($t \to b \ell N) \sim 2 
\times 10^{-4}$. However, it drops dramatically as soon as systematic 
uncertainties are included, due to the very small $S/B\sim 0.41\%$.
The asymmetry between the signal and the background in the BDT variable (see 
Fig.~\ref{fig:LHCanalysis2}) can however be used to better control the 
systematic uncertainties. 
To this aim, we  slide the bins from left to right in the corresponding 
distribution, and construct the variable $A = (N_{right} - N_{left})/(N_{right} 
+ N_{left})$, where $N_{left}$ ($N_{right}$) denotes the number of events to the 
left (right) of a chosen bin including the bin itself.
Systematic 
uncertainties cancel in this ratio. 
The number of standard deviations between the asymmetry in the 
signal + background and the SM alone is shown in Fig.~\ref{fig:finalTOP} as a 
function of BR($t \to b \ell N$). It can be seen that branching ratios as low 
as $\sim 2\times 10^{-4}$ can be tested at the LHC in the long run. Scaling 
with the larger $t\overline{t}$ cross section, we also estimate the reach at 
$\sqrt{s}=27$~TeV with $\mathcal{L}=10$~ab$^{-1}$, that improves by a factor of 
$\sim 4$.
The projected 
95\% C.L. bounds on the various couplings entering this rare decay, as described 
in Eq.~\ref{eq:tTOblN}, are shown in Tabs.~\ref{tab:bounds1} 
and~\ref{tab:bounds3} for the HL-LHC analysis. 

In summary, we find that the BDT combined with the asymmetry variable helps in 
constraining these EFT couplings. More studies are required in this direction 
and a control over systematics is warranted.

\section{Conclusions}
\label{sec:conclusions}
%
By using LHC searches for light leptons or taus with missing energy, monojet 
analyses, and measurements of different pion and tau decay widths, we have singled 
out the most constrained directions in the $\nu$SMEFT and, as a consequence, 
those others in which new physics can be more likely hiding. (Our results are 
valid for a cutoff above the EW scale provided $N$ is the RH component 
of the SM Dirac neutrino or a new Majorana particle with $0.01~\mathrm{GeV}\lesssim m_N\lesssim 0.1$ GeV.)

In the first category, our limits range from $\sim 380$ ($1400$) GeV to $\sim 
3$ ($11$) TeV for couplings of order $\sim 1$ ($4\pi$); corresponding to 
leptonic operators with taus and first generation quark-lepton operators,  
respectively. In the category of unconstrained operators, 
we find interactions involving heavier 
quarks, as well as operators that could be better probed at lepton facilities. 
However, 
the operators giving genuinely new signals are those providing new top decays. 
In particular, the operators $\mathcal{O}_{duNe}$, $\mathcal{O}_{ldqN}$, 
$\mathcal{O}_{lNqd}$ and $\mathcal{O}_{quNl}$ trigger the process $t\to 
b\ell+\text{inv}$, with the missing energy and the lepton not reconstructing a 
$W$ boson. This not yet explored process is the only sensible signature of 
these operators. We have worked out a BDT analysis sensitive to this decay 
channel in semileptonic $t\overline{t}$ production relying mostly on the 
invariant mass of the lepton and the missing energy; obtained 
upon requiring the top decay products to reconstruct the top mass.

The BDT variable turns out to be sensibly shifted in the signal with 
respect to the background. In order to minimise systematic uncertainties, we 
have based the statistical analysis on an asymmetry built out of the BDT 
variable. Considering one single lepton family, we have shown that, at the 
LHC with $\sqrt{s} = 14$ ($27$) TeV and $3$ ($10$) ab$^{-1}$ of collected 
luminosity, branching ratios of the top quark into this exotic decay mode as 
small as $\sim 2\times 10^{-4}$ ($\sim 5\times 10^{-5}$) can be probed at the 
95\,\% C.L. This translates into a prospective lower bound on $\Lambda\sim 330$ 
(460)~GeV for $\alpha\sim 1$; $\Lambda\sim 1.2$ (1.6)~TeV for 
$\alpha\sim 4\pi$. These numbers 
can rise up to $\Lambda\sim 1.8$ ($2.5$) TeV in the strongly interacting 
limit if both electrons and muons as well as three RH neutrinos are present. 

Finally, let us emphasise that the aforementioned results are also useful to 
constrain certain operators in the usual SMEFT. Indeed, although SMEFT 
operators contributing to $t\to b\ell+\text{inv}$ typically also induce  
quark-charged lepton interactions (because LH neutrinos share $SU(2)_L$ doublet with the charged
leptons; and in this respect the $\nu$SMEFT is qualitatively different), some 
directions can be mostly constrained upon exploiting the new top decay we have just studied. For example, let us expand the operator
\begin{align}
\nonumber
 \mathcal{O} &= (\overline{l_{iL}}\gamma^\mu l_{iL}) 
(\overline{q_{3L}}\gamma_\mu q_{3L}) + (\overline{l_{iL}}\gamma^\mu\sigma^I 
l_{iL}) (\overline{q_{3L}}\gamma_\mu \sigma_I
q_{3L})\\
&=(\overline{\nu_{iL}}\gamma^\mu\nu_{iL})(\overline{t_L}\gamma_\mu t_L) + 
(\overline{\ell_{iL}}\gamma^\mu \ell_{iL})(\overline{b_L}\gamma_\mu 
b_L)+\bigg[\boldsymbol{(\overline{\ell_{iL}}\gamma^\mu 
\nu_{iL})(\overline{t_L}\gamma_\mu b_L)}+\text{h.c.}\bigg]~.
\end{align}
%
The non-boldface interactions are very hard to probe at hadron colliders. At 
lepton facilities, one could study the $e^+e^	-\to b\overline{b}$, but any 
departure from the SM could be attributed to \textit{e.g.} RRRR operators, 
unless the rare top decay is also tagged.

\bigskip
\noindent
\textbf{Note added.} 
During the final stages of this work, Ref.~\cite{Bischer:2019ttk} appeared, 
in which the bounds on some operators coming from neutrino experiments 
are derived.

\section*{Acknowledgements}
\noindent
We would like to thank Gurpreet Chahal, Xiao-Dong Ma, 
Arcadi Santamaria and Jakub Scholtz for helpful discussions. 
JA is supported by the Spanish MINECO under grant FPA2014-54459-P. 
SB is supported by a Durham Junior Research Fellowship COFUNDed 
between Durham University and the European Union under grant agreement 
number 609412.
MC is supported by the Spanish MINECO under the Juan de la Cierva programme and by
the Royal Society under the Newton International Fellowship programme. 
AT acknowledges funding from the European Union's Horizon 2020 research and innovation programme under the Marie 
Sk\l{}odowska-Curie grant agreement No 674896 (ITN Elusives). 
AT would also like to thank CERN, where part of this work was carried out, 
for kind hospitality.

\bibliography{vSMEFT_v2}

\end{document}